\documentclass[%
reprint,
superscriptaddress,
aps,
prl,
portrait,
]{revtex4-2}
\usepackage{pdfpages}
\usepackage{blindtext}
\usepackage{amsmath}
\usepackage{bm}
\usepackage{xcolor}
\usepackage{soul}
\usepackage{graphicx}
\usepackage{subfig}
\graphicspath{ {./figures/} }

\makeatletter
\AtBeginDocument{\let\LS@rot\@undefined}
\makeatother

\begin{document}


\title{Resonantly-driven nanopores can serve as nanopumps}

\author{Aaron D. Ratschow}
\affiliation{
 Institute for Nano- and Microfluidics, TU Darmstadt,\\
 Alarich-Weiss-Stra{\ss}e 10, D-64237 Darmstadt, Germany
}
\author{Doyel Pandey}
\affiliation{
 Department of Mathematics, Indian Institute of Technology Kharagpur,\\
 Kharagpur, West Bengal, India - 721302
}
\author{Benno Liebchen}
\email{benno.liebchen@pkm.tu-darmstadt.de}
\affiliation{
 Theory of Soft Matter, Department of Physics, TU Darmstadt,\\
 Hochschulstra{\ss}e 12, D-64289 Darmstadt, Germany
}
\author{Somnath Bhattacharyya}
\affiliation{
 Department of Mathematics, Indian Institute of Technology Kharagpur,\\
 Kharagpur, West Bengal, India - 721302
}
\author{Steffen Hardt}
\email{hardt@nmf.tu-darmstadt.de}
\affiliation{
 Institute for Nano- and Microfluidics, TU Darmstadt,\\
 Alarich-Weiss-Stra{\ss}e 10, D-64237 Darmstadt, Germany
}

\date{July 19, 2022}

\begin{abstract}
Inducing transport in electrolyte-filled nanopores with dc fields has led to influential applications ranging from nanosensors to DNA sequencing. Here we use the Poisson-Nernst-Planck and Navier-Stokes equations to show that unbiased ac fields can induce comparable directional flows in gated conical nanopores. This flow exclusively occurs at intermediate driving frequencies and hinges on the resonance of two competing timescales, representing space charge development at the ends and in the interior of the pore. We summarize the physics of resonant nanopumping in an analytical model that reproduces the results of numerical simulations. Our findings provide a generic route towards real-time controllable flow patterns, which might find applications in controlling the translocation of particles such as small molecules or nanocolloids.


\end{abstract}

\maketitle

\paragraph*{Introduction --}
In the past few years, nanochannels and nanopores have been the subject of rapidly intensifying research activities. They have found applications as nanosensors \cite{Zhang.2021,Rahman.2021}, in DNA sequencing \cite{Wanunu.2012,Feng.2015}, for liquid and gas-phase separation processes \cite{Wang.2020}, for water desalination by reverse osmosis \cite{Liu.2021}, or for power generation from salinity gradients \cite{Macha.2019}. 
For many applications of electrolyte-filled nanochannels and nanopores, it is desirable to control the transport processes inside the channels or pores. For such purposes, control schemes based on a gate voltage, often applied via a gate electrode, have been developed. By tuning the gate voltage, the electrostatic potential at the channel or pore walls can be modulated. In turn, the charge in the diffuse part of the electric double layer (EDL) can be controlled, which is especially interesting in the case of overlapping EDLs. In experiments it was demonstrated that corresponding schemes enable controlling the transport properties of nanopores. Examples include the control of water permeation \cite{Zhou.2018} and DNA translocation \cite{Yen.2012,Liu.2016,Xue.2018}, and more generally, the control of electromigration and diffusion fluxes through nanopores \cite{Karnik.2005,Fuest.2015,Fuest.2017,Cheng.2018,Ren.2018,Xue.2021} which can even be rectified \cite{Kalman.2009,Guan.2011,Laucirica.2019,Li.2021}. As for the latter, employing nanopores with conical or converging geometries has proven particularly useful \cite{Apel.2001,Poggioli.2019}. For example, when filled with an electrolyte solution, their electric resistance can become direction-dependent, giving such pores current-rectifying properties \cite{Wei.1997,Siwy.2002,Siwy.2002b}. This results in asymmetric current-voltage I(V) curves, such that the magnitude of the resulting current depends on the sign of the applied dc voltage \cite{Jin.2010}.  
Notably, to date, most existing transport control and rectification schemes in nanopores employ a dc driving (or a slow ac-driving), based on a constant applied voltage, such that the transport properties are essentially controlled by the equilibrium or steady-state properties of the system. Compared to that, dynamic aspects have received little attention. \\
In the present work, we explore the transport properties of conical nanopores under the influence of ac driving, i.e. for gate-electrodes energized with frequencies high enough to keep the ion cloud in the pore persistently away from its equilibrium configuration. By combining simulations of the Poisson-Nernst-Planck (PNP) and Navier-Stokes (NS) equations and analytical modeling, we show that an ac driving induces a novel and subtle transport mechanism in which the pore serves as a nanopump creating a net fluid flow – even in the absence of a direction-dependent flow resistance. Remarkably, the pumping neither occurs for very slow or very fast driving, but the maximum flow rate is reached at a characteristic resonance frequency which can be determined by the intrinsic relaxation rate of the system in response to a step change in voltage. We call this mechanism resonant nanopumping. We explain the mechanism using an analytical model which predicts the multi-dimensional parametric dependency of the flow rate, in quantitative agreement with our detailed simulations.
Our results unveil a novel nanopumping principle which offers the ac driving frequency as a parameter to control the transport through a pore. 
In addition, we note that the present work links the physics of nanopores to Brownian ratchets \cite{Reimann.2002,Hanggi.2009,Denisov.2014} and could induce a transfer of ideas between these research fields. 
\begin{figure*}
  \includegraphics[width=0.99\textwidth]{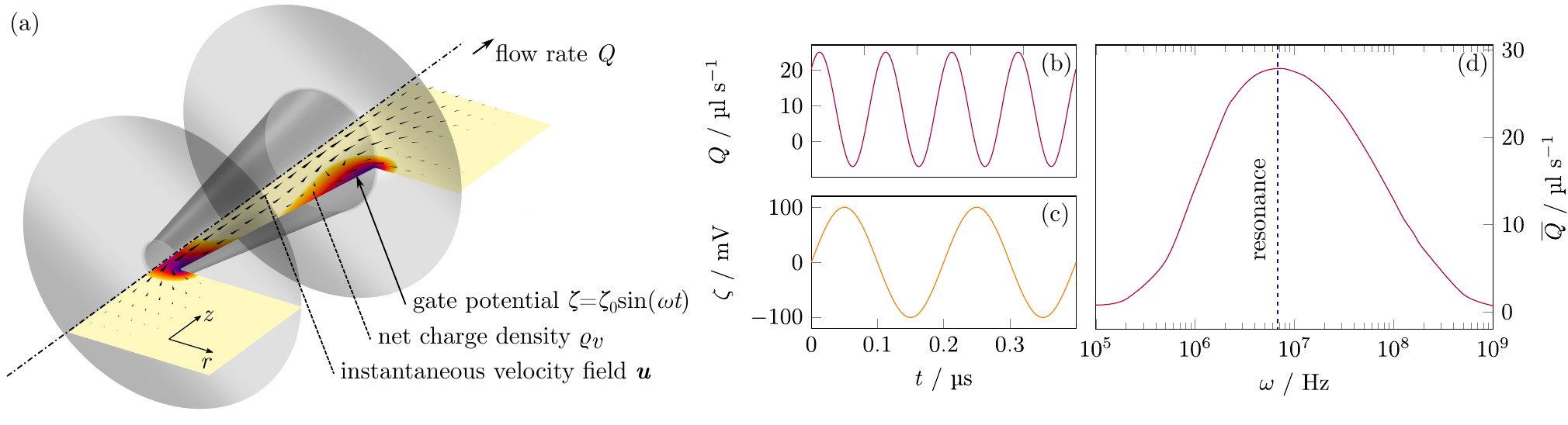}
  \caption{Resonant nanopumping: (a) Schematic illustration of the conical nanopore connecting two reservoirs (indicated by grey discs), the flow field (arrows) and the charge density (colors, ranging from low (yellow) to high (purple) charge density). 
  (b)-(d)
Applying an unbiased ac gate potential $\zeta$ with a frequency $\omega$ (c) to the pore wall 
leads to a flow $Q$ that oscillates with 
$2\omega$ and is biased towards the wide end of the pore (b). The magnitude of the mean flow $\overline{Q}$ 
strongly depends on $\omega$, with a distinct resonant behavior (d).}
\end{figure*}
\paragraph*{Setup --}
We consider a conical nanopore with length $L=100$ nm, a radius at the narrow end $r_n=7.5$ nm and an opening angle of $\alpha = 10^\circ$, which connects two reservoirs filled with an aqueous, monovalent, binary electrolyte solution (Fig. 1). 
To avoid singularities, the corners at the two ends of the pore are blunted with a radius of 1 nm. Both reservoirs feature the same prescribed pressure and ion concentrations $c_0$, and their boundaries far from the pore (cropped in Fig. 1a) are grounded. 
Ion concentrations such that the Debye length $\lambda_D$ is smaller than $r_n$ are chosen. 
The pore wall is assumed to exhibit no native zeta potential and the gate potential at the wall $\zeta(t)$ can be varied. The pore and reservoir walls (grey in Fig. 1a) are considered non-slipping and impermeable for ions. 

\paragraph*{Flow field --}
Let us now explore the flow through the pore when applying an ac gate potential $\zeta(t)=\zeta_{\mathrm{0}}\mathrm{sin}(\omega t)$ of angular frequency $\omega$. To do this, we numerically solve the fully coupled set of PNP and NS equations on this axisymmetric two-dimensional geometry using the finite-element method (see Supplemental Material (SM) for details). We monitor the flow rate $Q= \int_{\Gamma}u_zdA$ (Fig. 1b and d), where $u_z$ is the axial velocity and $\Gamma$ is any cross-sectional plane within the pore.
\\When $\omega$ tends towards zero, we observe that the time-averaged flow vanishes $\overline{Q} \to 0$ (Fig. 1d). This is plausible, as for slow driving anions and cations alternately enter and leave the pore via both openings and form fully developed EDLs along the entire pore wall over the full driving cycle (Fig. 2, left panel). 
Consequently, the electric body force within the equilibrium EDLs is counteracted by the electrohydrostatic (osmotic) pressure gradient and the net force on the fluid vanishes at all times \cite{Levich.1962}. 
Similarly, when $\omega$ tends towards infinity, we observe a rapidly oscillating but very weak flow $Q$, and $\overline{Q}$ again tends towards zero (Fig. 1d). 
Reason is that 
for very fast driving, ions are alternately driven towards and away from the pore wall to screen the gate potential, but do not have enough time to migrate into  the pore. Accordingly, we observe small ($\ll \lambda_D$) charged regions at the corners, whereas the liquid around the pore center remains uncharged (Fig. 2, right panel). In this case, charged regions are virtually absent and the flow is very weak. 
Remarkably, however, for 
intermediate $\omega$, we observe 
a significant flow. It oscillates with twice the driving frequency and shows a strong bias, such that the net flow $\overline{Q}$ points from the narrow to the wide end of the pore (Fig. 1c).
The emergence of such a flow might surprise, since there are no apparent driving gradients, especially pressure gradients, induced by the boundary conditions. 
To understand how this flow comes about, note that after each sign change of the gate potential, new EDLs start to form at both pore ends and spread towards the pore center. This process gets interrupted by the subsequent sign change of the gate potential such that the EDLs follow the gate potential with a delay, causing charge density gradients  along the pore that result in significant axial electric fields at the pore ends. The middle panel of Fig. 2 even shows the simultaneous presence of oppositely charged regions. 
These axial electric fields exert a body force $\rho_v \bm{E}$ on the charged liquid at the corners, that oscillates with twice the driving frequency $2\omega$, since both the charge density $\rho_v$ and the electric field $\bm{E}$ oscillate with $\omega$. Because of the conical pore shape, these body forces push fluid into the pore with different strength at the two ends, leading to the observed bias which we will further discuss and quantify below. 
\begin{figure}
  \includegraphics[width=0.95\columnwidth]{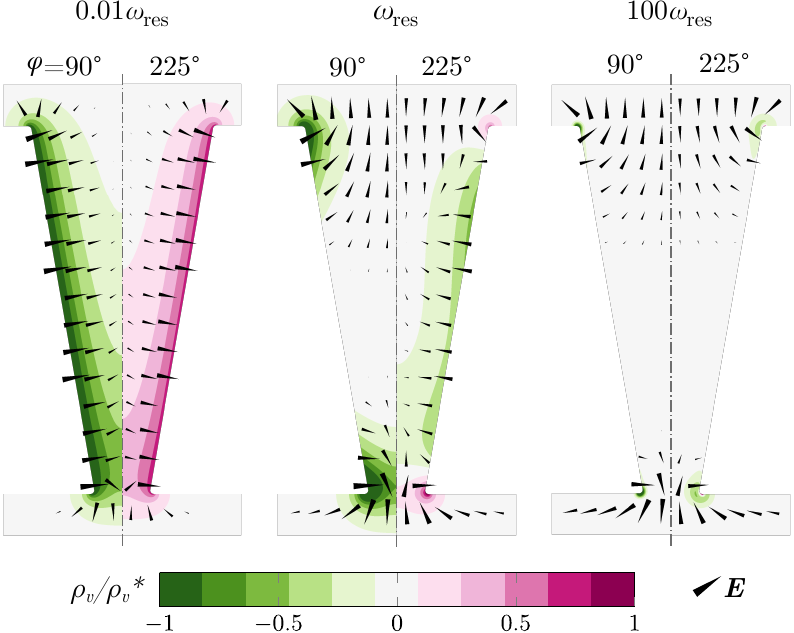}
  \caption{Net charge density $\rho_v$ and electric field vectors $\bm{E}$ within the nanopore for slow, resonant, and fast driving with $\zeta=\zeta_0\mathrm{sin}(\omega t)$. The scales $\rho_v^*$ are 250, 80, and 20 kCm$^{-3}$ respectively, $\bm{E}$ is scaled for optimized visualization. Splits show instantaneous fields at different phase angles $\phi=\omega t$. }
\end{figure}

\paragraph*{Step response and characteristic timescales --}
To understand the complex interplay of space charge and electric field, we 
explore the response of the system (
assumed to be in equilibrium at $t=0$) to a step-like gate potential $\zeta(t)=\Theta(t) \zeta_0$ where $\Theta(t)=1$  for $t\geq 0$ and zero elsewise (method of step response \cite{Leigh.2012}). 
Such a gate potential causes a fast increase of the flow rate towards a distinct maximum, which is followed by a comparatively slow relaxation towards the new  equilibrium state (see Fig. 4a). That is, the system possesses two competing inherent timescales: 
a short one corresponding to the time needed for developing EDLs at the corners, and a long timescale corresponding to the time needed for the ions to migrate into the pore and develop an EDL throughout.
Within the latter timescale, the axial electric field vanishes, ultimately leading to an electric field that points in wall-normal direction in all space-charge regions (Fig. 3, left splits). 
\begin{figure}
  \includegraphics[width=0.95\columnwidth]{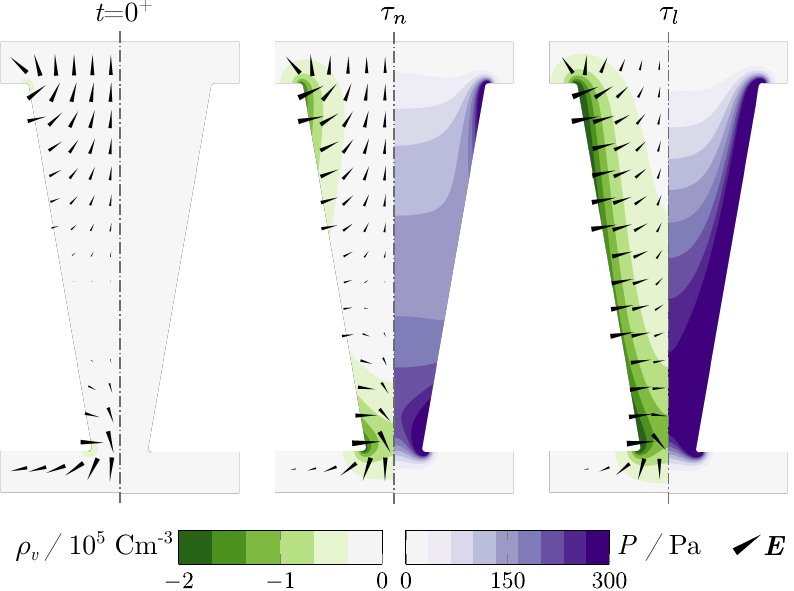}
  \caption{Net charge density $\rho_v$ with electric field vectors $\bm{E}$ (left split) and pressure $P$ (right split) right after the gate potential step and after the short and long timescales. An axial pressure gradient quickly develops in response to electric body forces acting at the corners. For long times, equilibrium EDLs form, and the wall-normal electric body force and the pressure gradient cancel.}
\end{figure}
When one pore end is viewed independently, the only lengthscale that the electric field can depend on is the pore diameter. 
Thus, after switching on the gate potential at $t=0$, there is a significant axial electric field which essentially persists over that lengthscale (Fig. 3, left and middle panels). Since the interaction of charges with this field drives the flow, we identify the short timescale as the time of diffusive transport of counterions into the pore over a length of one pore diameter to form EDLs at the corners of the pore (Fig. 3, middle panel). The transport of co-ions in the opposite direction proceeds in a similar manner. 
Notably, although $r_i$ can be $\sim \lambda_D$ in this work, we find that the charging timescale for a pore end largely follows the simple expression 
\begin{equation}
    \tau_i=4\frac{\lambda_D r_i}{D}, \: i=n,w,
\end{equation}
obtained from the transmission line model (TLM) for $\lambda_D \ll r_i$ (see \cite{Mirzadeh.2014} and SM). Here $D$ is the ion diffusivity and the subscripts $n$ and $w$ denote the narrow and wide ends, respectively.
\\
The long timescale is associated with the filling of the complete pore with counterions, leading to fully evolved EDLs and to a cancellation of the axial electric field. It is the timescale of diffusive transport of ions from both pore ends over a distance of roughly $L/2$, which is again derived from the TLM as (see SM) 
\begin{equation}
    \tau_l=\frac{1}{4}\frac{\lambda_D}{\overline{r}}\frac{L^2}{D}, 
\end{equation}
where $\overline{r}=1/2\left(r_n+r_w \right)$ is the mean-radius of the conical pore.
 
\paragraph*{Delay effects drive the nanopump --}
We now further resolve the crucial role of delay effects for the operation of the nanopump. To this end, note first that for the step response the axial electric body force $\rho_v E_{z}$ only acts in the time window when $\rho_v$ has already developed and $E_{z}$ has not yet vanished, i.e. roughly for $\tau \in [\tau_i,\tau_l]$. 
Within this time window, the body force pushes liquid into the pore from both ends, irrespective of the sign of $\zeta$. The conical pore shape breaks the reflection symmetry with respect to $z$, making the force contributions from both ends unequal. As a result of these localized forces, an axial pressure gradient develops (Fig. 3, right split) and liquid flows from the narrow to the wide end. 
\\Accordingly, when driving the pore with an ac voltage with a frequency large enough to cause significant delay effects, but small enough to allow EDLs to partially propagate into the pore, we observe unequal oscillating axial body forces at both ends of the pore that drive liquid through it.


\paragraph*{Analytical model --}
To quantitatively predict the flow through the nanopore, we translate the localized axial body force contributions into effective pressures at both pore ends, matching the pressure field emerging in reaction to these forces (see Fig. 3, middle panel). We then solve Reynolds' lubrication equation to predict the resulting flow.\\
The effective local pressure at either pore end can be estimated by area-averaging the axial electric body force, assuming it is fully present only in an annular region of width $\lambda_D$ – corresponding to the EDL – with an axial extension comparable to the pore radius. For the step potential, this yields a time dependent pressure at a pore end $i$ of 
 \begin{equation}
     P_i(t)=\frac{r_i^2-(r_i-\lambda_D)^2}{r_i} \:\: \Tilde{\rho}_v \! \left(1-e^{-t/\tau_i} \right) \: \Tilde{E}_{z,i}e^{-t/\tau_l},\label{eq:Pi_SR}
 \end{equation}
 where we use tildes to denote characteristic scales and assume exponential decay on each timescale, due to the capacitor-like charging of the EDL \cite{Janssen.2021}. Clearly, this pressure is higher at the narrow end of the pore. 
 We estimate the characteristic volume charge via the Debye-H\"uckel approximation as $\Tilde{\rho}_v \sim 2\zeta_0c_{0}F^2/(eRT)$, with temperature $T$ and the Faraday and gas constants $F$ and $R$, respectively. (See SM for details).
Assuming, again, that the electric field at both pore ends first emerges on a lengthscale comparable to the pore diameter (Fig. 3) yields 
$\Tilde{E}_{z,i} \sim \zeta_0/(2r_i)$. 
The pressures in eqn. \ref{eq:Pi_SR} are proportional to the product of the step responses of volume charge and axial electric field. Thus, we find the time-dependent pressures in response to an ac gate potential $\zeta(t)=\zeta_0\mathrm{sin}(\omega t)$ from the step responses \cite{Berns.2021} and again area-average their product (see SM): 

 \begin{widetext}
\begin{equation}
     P_i(t,\omega)=K_i\zeta_0^2\frac{\omega \tau_l}{\sqrt{1+\omega^2\tau_i^2} \sqrt{1+\omega^2\tau_l^2}}\frac{1}{2} \bigg\{ \mathrm{cos}\left[ \mathrm{atan}(\omega\tau_i)- \mathrm{atan}(\frac{1}{\omega\tau_l}) \right]- \mathrm{cos}\left[2\omega t + \mathrm{atan}(\omega\tau_i)+ \mathrm{atan}(\frac{1}{\omega\tau_l}) \right] \bigg\},
\end{equation}
 \end{widetext}
\begin{equation}
    K_i=c_{0}\frac{F^2}{eRT}\frac{r_i^2-(r_i-\lambda_D)^2}{r_i^2}, \: i=n,w.
\end{equation}

The first term in braces is a frequency-dependent constant offset and the second term describes a harmonic oscillation with twice the driving frequency and a frequency-dependent phase shift. To translate the effective pressures at the pore ends into a flow rate, we analytically solve Reynolds' lubrication equation for a pressure-driven flow in a conical pore, yielding (SM) 
\begin{equation}
    Q=-\frac{3}{8}\frac{\pi}{\mu}\frac{\frac{r_w-r_n}{L}r_n^3}{1-\frac{r_n^3}{r_n^3+r_w-r_n}}(P_w-P_n),
    \label{Qpred}
\end{equation}
where $\mu$ is the dynamic viscosity of the liquid. 
The resonance frequency $\omega_{\mathrm{res}}$ readily follows from this expression by (numerically) solving $\partial \overline{Q}/\partial \omega=0$ for $\omega > 0$. 
Overall, eqns. 4-6 (or 3 and 6) represent a full analytical model of the flow rate through the conical nanopore for a harmonic (or step-like) gate potential, which does not contain any free parameters. 
The model is expected to make valid
predictions if $\zeta_0 \ll RT⁄F$ and $\lambda_D<r_i \ll L$.

\begin{figure}
	\includegraphics[width=0.85\columnwidth]{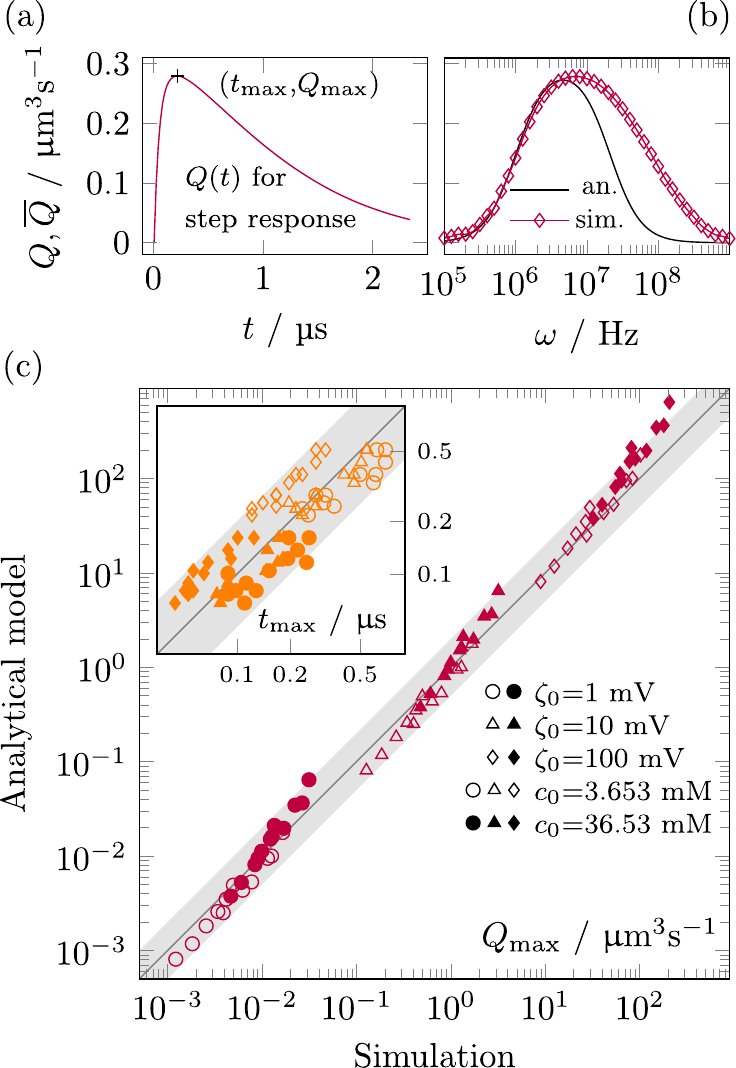}
	\caption{
		(a) Step response of the system. (b) Mean flow rate $\overline{Q}$ over ac driving frequency for the analytical model and simulations; symbols denote time-domain simulations of the governing equations. (c) Scatter plots comparing the analytical and simulation results over a large domain of the parameter space for step responses, using the maximum flow rate $Q_\text{max}$ and time of maximum flow rate $t_\text{max}$ (inset), with shaded factor-two error areas. 
	}\label{fig4}
\end{figure}

\paragraph{Comparison between model and numerical simulations --}
Let us now systematically test the model by comparing it to our numerical simulations. 
To do this, we compare the maximum flow rate $Q_{max}$ and the time where this maximum occurs $t_{max}$ for the
step response simulations (Fig. 4a) in a 5-dimensional parameter space with $\zeta_0\in \{1,10,100 \}$ mV, $c_0\in \{3.6532,36.532 \}$ mol m${}^{-3}$,   $\:r_n\in \{7.5,15 \}$ nm, $L\in \{50,100,200 \}$ nm, and $\alpha\in \{5,10,20 \}^\circ$.
The result shows a good agreement over the entire domain of the parameter space (Fig. 4c), even for $\zeta_0=100\:\mathrm{mV} > 25\: \mathrm{mV}\approx RT/F$, which is beyond the regime where our assumptions are justified.
\\ Note that due to the linearity of the underlying charging processes, the comparison for the step response is largely representative also for the harmonic gate potential.
As exemplarily shown in Fig. 4b,  the model correctly predicts the mean flow rate up to the resonance frequency.
Noticeable deviations at larger frequencies can be attributed to the fact that the model assumes driving axial fields at the pore ends only governed by the delay of screening charge at the pore center (timescale $\tau_l$).
However, at high frequencies, the pore center remains uncharged at all times and otherwise negligible variations due to the local dynamics of charge generation at the corners (timescale $\tau_i$) become dominant. This introduces nonlinearities that are not accounted for in the model.

\paragraph{Experimental feasibility --}
Several studies have proven manufacturability of gated nanopores \cite{Karnik.2005,Joshi.2010} including conical shapes of similar dimensions as previously discussed \cite{Kalman.2009, Nam.2009}. The proposed mode of resonant ac driving is thus practically realizable. Moreover, it is not limited to the pore dimensions considered above, since the fundamental physical mechanism applies more broadly.

\paragraph{Conclusions and outlook --}
To conclude, we have demonstrated that electrolyte-filled conical nanopores with gate electrodes create a net flow when an ac voltage is applied to the gate. We have explained the emergence of a net flow as a resonance phenomenon that is based on two competing timescales. Achievable flow velocities range from millimeters to centimeters per second, which is comparable to electroosmotic flow in nanochannels and other nanopumping principles [26-32]. Resonant nanopumping could allow real-time flow regulation in individually addressable nanopores, where the voltage amplitude and driving frequency serve as parameters to control the magnitude and temporal structure of the flow field.  Generally, gated transport through nanopores under ac driving could open up superior possibilities for selective mass transfer, like in the case of macromolecules or nanoparticles. As an example, the principle could prove beneficial for controlling DNA translocation in sequencing applications. Last but not least, the physical mechanisms related to flow induced by transient electric double layers uncovered in this work could find applications with a number of different electrode structures beyond conical shapes.

\begin{acknowledgments}
We wish to acknowledge the help provided by Maximilian T. Sch\"ur in preparation of the figures. 
\end{acknowledgments}
S. H. and B. L. proposed the work, A. D. R. and D. P. carried out the simulations, A. D. R. developed the theoretical framework and the analytical model, A. D. R., B. L. and S. H. contributed to the interpretation of the results,  A. D. R., B. L. and S. H. prepared the manuscript, and S. H. and S. B. supervised the work.

\bibliographystyle{apsrev4-1} 

%

\clearpage


\section*{Supplementary material}

# RESONANTLY-DRIVEN NANOPORES CAN SERVE AS NANOPUMPS

## -Supplemental Material-

Aaron D. Ratschow,[1] Doyel Pandey,[2] Benno Liebchen,[3, *] Somnath Bhattacharyya,[2] and Steffen Hardt[1, †]

[1]*Institute for Nano- and Microfluidics, TU Darmstadt,*
*Alarich-Weiss-Straße 10, D-64237 Darmstadt, Germany*
[2]*Department of Mathematics, Indian Institute of Technology Kharagpur,*
*Kharagpur, West Bengal, India - 721302*
[3]*Theory of Soft Matter, Department of Physics, TU Darmstadt,*
*Hochschulstraße 12, D-64289 Darmstadt, Germany*
(Dated: July 19, 2022)

The Supplemental Material consists of two main sections. §1 explains the numerical setup and procedure. In §2, details about the analytical model and its derivation are given.

## §1: NUMERICAL METHOD

### §1.1: Governing equations and boundary conditions

All numerical results are obtained based on the coupled set of Poisson-Nernst-Planck and Navier-Stokes equations for an incompressible, Newtonian liquid, and a symmetric binary electrolyte solution of monovalent ions. The governing equations are solved in their dimensional form and read

$$\nabla^2 \Phi = -\frac{F(c_1 - c_2)}{\epsilon}, \tag{S1}$$

$$\frac{\partial c_i}{\partial t} + \nabla \cdot (\mathbf{J}_i) = 0, \ \mathbf{J}_i = \mathbf{u} c_i - D\left(\nabla c_i + \frac{z_i F}{RT} c_i \nabla \Phi\right), \ i = 1, 2, \tag{S2}$$

$$\rho\left[\frac{\partial \mathbf{u}}{\partial t} + (\mathbf{u} \cdot \nabla) \mathbf{u}\right] = -\nabla P + \mu \nabla^2 \mathbf{u} - F(c_1 - c_2)\nabla\Phi, \tag{S3}$$

$$\nabla \cdot \mathbf{u} = 0, \tag{S4}$$

where $-\nabla \Phi = \mathbf{E}$ is the electric field, $F(c_1 - c_2) = \rho_v$ is the charge density, $\epsilon$ is the dielectric permittivity, $z_{1,2} = \pm 1$ is the valence of ions, and $\rho$ is the mass density. These equations are solved on a two-dimensional, axisymmetric domain representing the conical nanopore connecting two reservoirs (compare Fig. S1). The boundary conditions (BCs) are: $\mathbf{u} = 0$, $\Phi = \zeta(t)$, and $\mathbf{n} \cdot \mathbf{J}_i = 0$ at the pore wall, $\mathbf{u} = 0$, $\mathbf{n} \cdot \nabla\Phi = 0$, and $\mathbf{n} \cdot \mathbf{J}_i = 0$ at the reservoir walls, $\mathbf{n} \cdot (-P\mathbf{I} + \mu(\nabla\mathbf{u} + (\nabla\mathbf{u})^T)) = 0$, $\Phi = 0$, and $c_i = c_0$ at the circular-arc reservoir boundaries as well as $\mathbf{n} \cdot \mathbf{u} = \mathbf{n} \cdot \nabla\Phi = \mathbf{n} \cdot \mathbf{J}_i = 0$ at the symmetry axis.

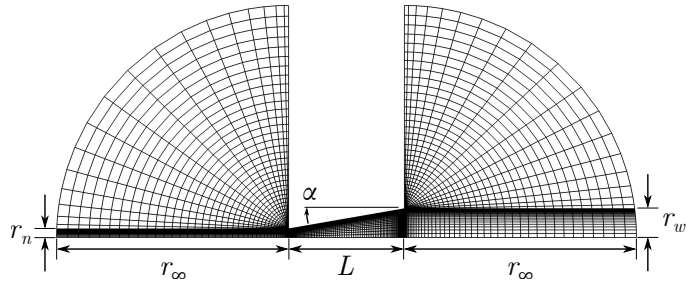

FIG. S1. Schematic representation of the computational domain and structured grid for $h = 1$ and $r_\infty/L = 2$. The reservoirs are approximated by quadrants of radius $r_\infty$. Note that all simulations subsequent to verification of the numerical scheme were carried out using a refined grid and larger reservoirs, $h = 2$ and $r_\infty/L = 5$ (not shown due to poor visibility of the grid structure).

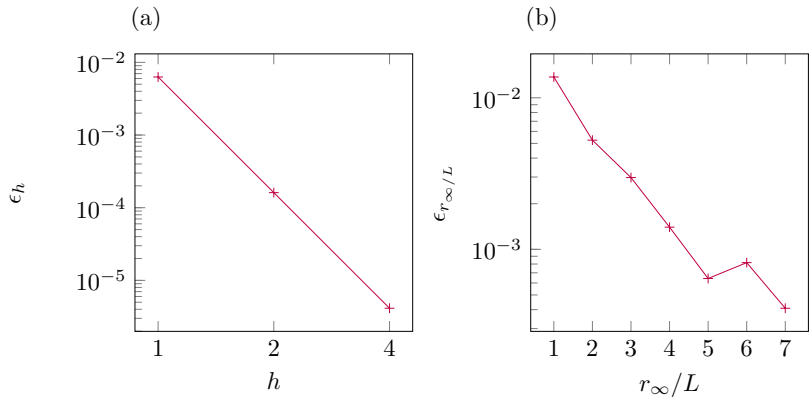

FIG. S2. (a) Relative error $\epsilon_h$ in volume flux $Q(\tau_n)$ obtained using the SR simulations for different levels of systematic grid refinement. The number of grid points per length in any direction is $\propto h$. The observed order of grid convergence is 5.29. (b) Relative error $\epsilon_{r_\infty/L}$ in volume flux $Q(\tau_n)$ obtained using the SR simulation for different reservoir sizes, expressed by $r_\infty/L$.

### §1.2: Implementation and verification

The geometry, governing equations, and BCs were implemented/solved based on the finite-element scheme of Comsol Multiphysics, version 6.0, using quadratic shape functions for $\Phi$ and $c_i$, and linear shape functions for $\mathbf{u}$ and $P$. The computational domain was discretized using a structured grid, as shown in Fig. S1. To avoid numerical instabilities in the time-integration scheme, the step function $\Theta(t)$ was smoothed across an interval of width $0.02\tau_n$ for all step-response (SR) simulations.

To ensure grid independence of the numerical solutions, the grid was systematically refined twice by inserting grid nodes halfway between existing nodes, which corresponds to refinement factors of 2 and 4, respectively, for the two refinements. In the SR simulations, the variable $Q(\tau_n)$ was chosen as a representative value for grid-quality assessment, and a Richardson extrapolation was performed to estimate the true, grid independent value $Q_R$. To ensure that the relative error due to discretization, $\epsilon_h = (Q_h(\tau_n) - Q_R)/Q_R$, is smaller than the desired numerical accuracy (for this work: 0.001), the grid $h = 2$ was chosen (see Fig. S2a). Similarly, finite-size effects with respect to the reservoirs were assessed by systematically varying the radius $r_\infty$, and monitoring the relative error $\epsilon_{r_\infty/L} = (Q_{r_\infty/L}(\tau_n) - \tilde{Q})/\tilde{Q}$, where $\tilde{Q} = Q_{r_\infty/L=10}(\tau_n)$ was used as an approximation to the solution at $r_\infty \to \infty$. Fig. S2b shows that $\epsilon_{r_\infty/L}$ becomes smaller than 0.001 for $r_\infty/L \geq 5$ and then randomly jumps due to the error being below the numerical accuracy. All subsequent simulations were carried out using $h = 2$ and $r_\infty/L = 5$.

## §2: ANALYTICAL MODELLING

### §2.1: Timescales

The resonant nanopumping phenomenon is governed by two timescales – a short one $\tau_i$ for the formation of electric double layers (EDLs) at the corners of either pore end $i$, and a long one $\tau_l$ for EDL formation in the entire pore. The derivation of both timescales is based on the transmission line model (TLM) of [S1]. The TLM states that for a pore of radius $h_p$, connected to an infinitely large reservoir, and in the limit of thin EDLs ($h_p \gg \lambda_D = \sqrt{\epsilon RT/(2c_0 F^2)}$), the charging front propagates into the pore on a timescale

$$\tau(x) = \frac{\lambda_D}{h_p} \frac{x^2}{D}, \tag{S5}$$

where $x$ is the propagation depth of the charging front into the pore. Notably, we adapt the scaling from the TLM to estimate both timescales.

The short timescale is associated with the formation of EDLs in the regions around the corners at the pore ends. This raises the challenge of finding a suitable definition of the relevant region around a pore end. Since the pore diameter is the only locally available length scale when a pore end is considered in isolation, we expect that the corresponding axial electric field emerges on this scale (see Fig. 3, left two panels). Therefore, we use the pore

diameter $2r_i$ as the relevant propagation depth $x$ and the pore radius $r_i$ as the radius $h_p$ in S5 to find the short timescale

$$\tau_i = 4\frac{\lambda_D r_i}{D}, \ i = n, w. \tag{S6}$$

For the long timescale, associated with EDL formation in the entire pore, we approximate the conical shape of the pore by a cylinder whose radius is obtained as the average of $r_n$ and $r_w$: $\bar{r} = 1/2(r_n + r_w)$. Since charging fronts propagate into the pore from both reservoirs simultaneously, the EDL formation can be considered complete when both fronts meet in the middle of the pore at length $L/2$. We thus use S5 with a propagation depth $x$ of $L/2$ and a radius $h_p$ of $\bar{r}$ to estimate the long timescale

$$\tau_l = \frac{1}{4}\frac{\lambda_D}{\bar{r}}\frac{L^2}{D}. \tag{S7}$$

### §2.2: Modeling effective pressures

As described in the main text, we assume that the space charge around the corners emerges on the short timescale $\tau_i$, while the relevant axial electric fields decay over the long timescale $\tau_l$, indicating EDL formation in the entire pore. Motivated by the charging/discharging dynamics of a capacitor, we further assume that both processes exhibit an exponential time response when the system is subject to a step change in gate potential $\zeta(t) = \zeta_0 \Theta(t)$. Using $\tilde{\rho}_v$ and $\tilde{E}_{z,i}$ as characteristic scales for the final charge density and the initial axial electric field, respectively, the axial component of the electric body force density in proximity to a pore end $i$ can be expressed as

$$f_{z,i} = \tilde{\rho}_v(1 - e^{-t/\tau_i})\tilde{E}_{z,i}e^{-t/\tau_l}. \tag{S8}$$

To find the corresponding effective pressures $P_i$ at either pore end that match the pressure field caused by the electric body forces, we assume that the pressure is constant over the pore cross section and only changes in axial direction. Further, we assume that $f_{z,i}$ only acts within the EDLs of thickness $\lambda_D$. Additionally, the volume charge decays in axial direction into the pore. Therefore, the region where the force is created does not extend over the entire length $2r_i$, so we use $r_i$ as an effective depth for averaging the force. Then, area averaging the force contribution at a pore end over the cross section yields the pressure over time for the SR

$$P_i(t) = \frac{r_i^2 - (r_i - \lambda_D)^2}{r_i}f_{z,i} = \frac{r_i^2 - (r_i - \lambda_D)^2}{r_i}\tilde{\rho}_v(1 - e^{-t/\tau_i})\tilde{E}_{z,i}e^{-t/\tau_l}. \tag{S9}$$

The Poisson equation S1 does not exhibit an inherent length scale, so for the electric field we use the locally available length scale at a pore end, the pore diameter, to find the characteristic scale $\tilde{E}_{z,i} \sim \zeta_0/(2r_i)$. For the characteristic scale of the volume charge $\tilde{\rho}_v$, we resort to the Debye-Hückel (DH) approximation of an EDL over a flat wall, which yields the potential and the ion concentrations as

$$\Phi_{DH} = \zeta_0 e^{-\eta/\lambda_D}, \tag{S10}$$
$$c_{i,DH} = c_0 e^{-z_i F/(RT)\Phi_{DH}}, \tag{S11}$$

where $\eta$ is the distance from the wall of potential $\zeta_0$. The charge density at $\eta = \lambda_D$ is deemed an appropriate characteristic scale for the volume charge. At this position, the potential is $\Phi_{DH}(\lambda_D) = \zeta_0/e$, and the charge density becomes

$$\begin{aligned}\tilde{\rho}_v = \rho_v(\eta = \lambda_D) &= F\left[c_{1,DH}(\lambda_D) - c_{2,DH}(\lambda_D)\right] \\ &= Fc_0\left(e^{-F\zeta_0/(eRT)} - e^{+F\zeta_0/(eRT)}\right) \\ &= -Fc_0 2\sinh\left(\frac{F\zeta_0}{eRT}\right) \\ &\approx -Fc_0\frac{2}{e}\frac{F\zeta_0}{RT},\end{aligned} \tag{S12}$$

where the sinh has been linearized in accordance with the DH approximation that assumes $\zeta_0 \ll RT/F$. Note that because both the volume charge and the electric field scale $\propto \zeta_0$, the pressures scale $\propto \zeta_0^2$, and the charged zones at the pore ends are always pushing liquid into the pore, irrespective of the sign of $\zeta_0$.

In order to translate the model for $P_i(t)$ in S9 from the SR scenario to a scenario with harmonic wall potential, we acknowledge that the pressure is proportional to the product of the SRs of the volume charge and the axial electric field,

$$\mathcal{SR}(P_i) = A_i \mathcal{SR}(\rho_{v,i}) \mathcal{SR}(E_{z,i}), \tag{S13}$$

where $\mathcal{SR}(\xi)$ denotes the SR of a quantity $\xi$, and $A_i = r_i^2 - (r_i - \lambda_D)^2/r_i$. From this we obtain, when $\mathcal{H}(\xi)$ denotes the harmonic response of $\xi$,

$$\mathcal{H}(P_i) = A_i \mathcal{H}(\rho_{v,i}) \mathcal{H}(E_{z,i}). \tag{S14}$$

Here, we assume independent linear responses of $\rho_{v,i}$, and $E_{z,i}$ (the justification for this assumption is given in §2.4). For linear, time-invariant systems, the response to a sinusoidal input $\propto \sin(\omega t)$, the SR, and the impulse response (IR) are connected by the Laplace transform via

$$\mathcal{IR}(\xi) = d/dt\, \mathcal{SR}(\xi), \tag{S15}$$
$$\mathcal{H}(\xi) = |\mathcal{L}\{\mathcal{IR}(\xi)\}| \sin\left[\omega t + \arg\left(\mathcal{L}\{\mathcal{IR}(\xi)\}\right)\right], \tag{S16}$$

where $\mathcal{L}\{\xi(t)\} = \int_0^\infty \xi(t) e^{-i\omega t} dt$ is the Laplace transform with imaginary argument $i\omega$, $\omega \in \mathbb{R}$ [S2]. Using S14-S16, we find the harmonic responses of the axial electric field and the volume charge,

$$\mathcal{H}(E_{z,i}) = \tilde{E}_{z,i} \frac{\omega \tau_l}{\sqrt{1 + \omega^2 \tau_l^2}} \sin\left[\omega t - \operatorname{atan}\left(\frac{1}{\omega \tau_l}\right)\right], \tag{S17}$$

$$\mathcal{H}(\rho_{v,i}) = \tilde{\rho}_{v,i} \frac{1}{\sqrt{1 + \omega^2 \tau_i^2}} \sin\left[\omega t - \operatorname{atan}(\omega \tau_i)\right], \tag{S18}$$

and after area-averaging, multiplication, and some rearrangement, the harmonic response of the pressure at a pore end $i$:

$$P_i(t, \omega) = K_i \zeta_0^2 \frac{\omega \tau_l}{\sqrt{1+\omega^2 \tau_i^2}\sqrt{1+\omega^2 \tau_l^2}} \frac{1}{2}\left\{\cos\left[\operatorname{atan}(\omega \tau_i) - \operatorname{atan}(\tfrac{1}{\omega \tau_l})\right] - \cos\left[2\omega t + \operatorname{atan}(\omega \tau_i) + \operatorname{atan}(\tfrac{1}{\omega \tau_l})\right]\right\}, \tag{S19}$$

$$K_i = c_0 \frac{F^2}{eRT} \frac{r_i^2 - (r_i - \lambda_D)^2}{r_i^2},\; i = n, w.$$

### §2.3: Solving Reynold's lubrication equation for the conical nanopore

In low Reynolds number flows driven by a volume force (here the electric body force), the local contributions of viscous forces, pressure gradient and volume force must balance everywhere in the fluid [S3]. Because the axial electric body force in this problem is restricted to localized regions at the pore ends, the fluid outside of these regions effectively follows the laws of a pressure-driven flow, where the pressure arises in response to the localized axial forces. Thus we model the flow by applying the effective pressures from §2.2 at the ends of the pore and solving for the pressure-driven flow within.

For this purpose, we consider an axisymmetric conical pore and use cylindrical coordinates, so that the symmetry axis of the pore is the $z$-axis and the radius of the pore is given by the expression $a(z) = a_0 + bz$, where in this case $a_0 = r_n$ and $b = (r_w - r_n)/L$. The starting point are the stationary, incompressible, axisymmetric Navier-Stokes and continuity equations

$$\rho\left(u_r \frac{\partial u_r}{\partial r} + u_z \frac{\partial u_r}{\partial z}\right) = -\frac{\partial P}{\partial r} + \mu\left[\frac{\partial}{\partial r}\left(\frac{1}{r}\frac{\partial}{\partial r}(r u_r)\right) + \frac{\partial^2 u_r}{\partial z^2}\right], \tag{S20}$$

$$\rho\left(u_r \frac{\partial u_z}{\partial r} + u_z \frac{\partial u_z}{\partial z}\right) = -\frac{\partial P}{\partial z} + \mu\left[\frac{1}{r}\frac{\partial}{\partial r}\left(r \frac{\partial u_z}{\partial r}\right) + \frac{\partial^2 u_z}{\partial z^2}\right], \tag{S21}$$

$$\frac{1}{r}\frac{\partial}{\partial r}(r u_r) + \frac{\partial u_z}{\partial z} = 0. \tag{S22}$$

We nondimensionalize with $\tilde{r} = r/a_0$, $\tilde{z} = z/L$, $\tilde{u}_r = u_r/U$, $\tilde{u}_z = u_z \epsilon/U$, and $\tilde{P} = P\epsilon Re/(\rho U^2)$, where in this case $Re = \rho U a_0/\mu$ and $\epsilon = a_0/L$. The average velocity of a pressure-driven flow through a cylindrical pore $U =$

$\bar{r}^2(P_n - P_w)/(8\mu L)$ represents a convenient velocity scale. After considering the asymptotic limits of the lubrication approximations, $Re \to 0$ and $\epsilon \to 0$, we arrive at the lubrication equations for this case,

$$\frac{\partial P}{\partial r} = 0, \tag{S23}$$

$$\frac{\partial P}{\partial z} = \mu \frac{1}{r}\frac{\partial}{\partial r}\left(r \frac{\partial u_z}{\partial r}\right), \tag{S24}$$

$$\frac{1}{r}\frac{\partial}{\partial r}(ru_r) + \frac{\partial u_z}{\partial z} = 0, \tag{S25}$$

given here in their dimensional form. We then proceed by solving S24 with the BCs $u_z(a) = 0$ and $\partial u_z/\partial r|_{r=0} = 0$. This leads to

$$u_z = \frac{1}{4}\frac{a^2}{\mu}\frac{\partial P}{\partial z}\left(\frac{r^2}{a^2} - 1\right), \tag{S26}$$

where it should be stressed that $a$ is a function of $z$. In the next step, we integrate the continuity equation S25 over the cross-sectional area, $\int_0^a \int_0^{2\pi} rd\phi dr$, and obtain

$$\int_0^a \frac{\partial}{\partial r}(ru_r)dr + \int_0^a \frac{\partial u_z}{\partial z} r dr = 0. \tag{S27}$$

Because no liquid can penetrate the boundaries at $r = 0$ and $r = a(z)$, it is obvious that the first term is zero and hence, the second term also has to vanish. We insert S26 and integrate, which results in an ordinary differential equation (Reynold's equation) for the pressure:

$$\begin{aligned}\int_0^a \frac{\partial u_z}{\partial z} rdr &= \int_0^a \frac{1}{4\mu}\left[\frac{\partial}{\partial z}\left(\frac{\partial P}{\partial z}r^2\right) - \frac{\partial}{\partial z}\left(\frac{\partial P}{\partial z}a^2\right)\right]rdr \\ &= -\frac{1}{16\mu}\frac{\partial}{\partial z}\left(\frac{\partial P}{\partial z}a^4\right) = 0 \\ &\implies \frac{\partial}{\partial z}\left(\frac{\partial P}{\partial z}a^4\right) = 0.\end{aligned} \tag{S28}$$

To solve for the pressure gradient, we integrate this equation and apply the BCs $P(0) = P_n$ and $P(L) = P_w$, which yields

$$\frac{\partial P}{\partial z} = \frac{P_w - P_n}{a^4}\frac{3ba_0^3}{1 - \frac{a_0^3}{(a_0 + bL)^3}}. \tag{S29}$$

Finally, we find the flow rate by integrating the axial velocity over any given cross-section and obtain

$$\begin{aligned}Q &= \int_0^a \int_0^{2\pi} u_z r d\phi dr \\ &= -\frac{1}{8}\frac{\pi a^4}{\mu}\frac{\partial P}{\partial z} \\ &= -\frac{3}{8}\frac{\pi}{\mu}\frac{ba_0^3}{1 - \frac{a_0^3}{(a_0 + bL)^3}}(P_w - P_n),\end{aligned} \tag{S30}$$

which is identical to equation 6 in the main text when the definitions of $a_0$ and $b$ are inserted.

### §2.4: Justification of the linearity assumption

In the derivation of the analytical model we assume that the inherently non-linear system can be approximated by the product of two independent linear processes, namely the buildup of EDLs at the corners and the propagation of EDLs into the interior of the pore, which in turn leads to the attenuation of the axial electric field. Here we provide justification for this assumption by comparing the system's numerically computed SR for different wall potentials.

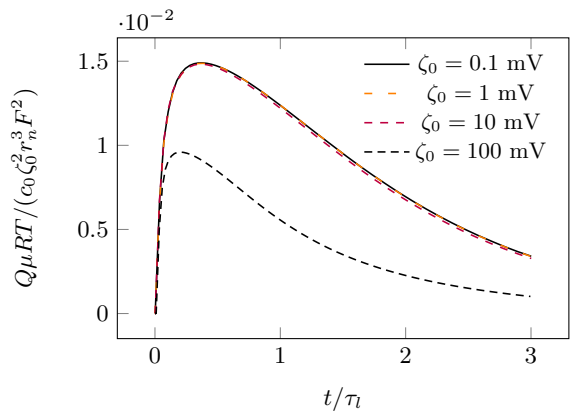

FIG. S3. Scaled SRs for different wall potentials $\zeta_0$, using the dimensions given in the main text and an ion concentration of $c_0 = 36.53$ mM. The curves collapse for $\zeta_0 \ll 25\, mV = RT/F$ with slight deviations when $\zeta_0$ approaches $RT/F$. Higher wall potentials introduce significant nonlinearities.

We scale the flow rate with $c_0 \zeta_0^2 r_n^3 F^2/(\mu RT)$ and the time with $\tau_l$ and expect the curves to collapse for different $\zeta_0$, since in the case of linear behavior, the characteristic timescales, $\tau_n$, $\tau_w$, and $\tau_l$, would only be functions of the pore geometry and the Debye length. Fig. S3 shows the SRs for $\zeta_0 \in \{0.1, 1, 10, 100\}$ mV. The curves corresponding to the three lowest wall potentials in fact collapse, with only slight deviations for $\zeta_0 = 10$ mV. Major deviations are observed for $\zeta_0 = 100$ mV. This result also holds for other geometries (not shown). The deviations indicate nonlinearities that can be attributed to wall potentials of the order of the thermal voltage $RT/F = 25$ mV or higher. In such cases, the Debye-Hückel approximation is violated and the system exhibits nonlinear behavior. Thus, we consider the assumption of linear processes during the transient charging of the nanopore appropriate when $\zeta_0 < RT/F$. Notably, the model still makes reasonable predictions for wall potentials up to 100 mV, as shown in Fig. 4c. However, in these cases, the model has a tendency to overpredict the flow, indicated by most datapoints lying above the diagonal of the scatterplot. This trend of $Q$ scaling more slowly with the wall potential than $\mathcal{O}\left(\zeta_0^2\right)$ for $\zeta_0 > RT/F$ is also visible in Fig. S3 and is expected to intensify for larger wall potentials.

---



\end{document}